\documentclass[conference]{IEEEtran}
\usepackage{graphicx}
\usepackage{setspace}
\usepackage{amsmath,empheq}
\usepackage{amssymb}
\usepackage{amsthm}
\usepackage{caption}
\usepackage{bm} 
\usepackage{cite}
\usepackage{float}
\usepackage[usenames,dvipsnames]{pstricks}
\usepackage{epsfig}
\usepackage{pst-grad} 
\usepackage{pst-plot} 
\usepackage{tabularx}
\usepackage{amsmath}
\usepackage[T1]{fontenc}
\newcommand{\lb} {\left}
\newcommand{\rb} {\right}
\newcommand{\nn} {\nonumber}
\usepackage{xcolor}
\usepackage{lipsum}
\usepackage{comment}

\newtheorem{theorem}{Theorem}

\allowdisplaybreaks

\begin{document}
 \onecolumn{\noindent © 2023 IEEE. Personal use of this material is permitted. Permission from IEEE must be obtained for all other uses, in any current or future media, including reprinting/republishing this material for advertising or promotional purposes, creating new collective works, for resale or redistribution to servers or lists, or reuse of any copyrighted component of this work in other works.}
 \twocolumn{
\title{Secrecy of Opportunistic User Scheduling in RIS-Aided Systems: A Comparison with NOMA Scheduling
\vspace{-0.6cm}}
\author{
    \IEEEauthorblockN{Burhan Wafai\IEEEauthorrefmark{1}, Sarbani Ghose\IEEEauthorrefmark{2}, Chinmoy Kundu\IEEEauthorrefmark{3}, 
    Ankit Dubey\IEEEauthorrefmark{4},
    and Mark F. Flanagan\IEEEauthorrefmark{5},
    }
    \IEEEauthorblockA{\IEEEauthorrefmark{1}\IEEEauthorrefmark{4}Department of EE, Indian Institute of Technology Jammu, Jammu \& Kashmir, India }
     \IEEEauthorblockA{\IEEEauthorrefmark{3}\IEEEauthorrefmark{5}School of Electrical and Electronic Engineering, University College Dublin, Belfield, Dublin 4, Ireland}
    \textrm{\{\IEEEauthorrefmark{1}burhan.wafai}, {\IEEEauthorrefmark{4}ankit.dubey}\}@iitjammu.ac.in,{\IEEEauthorrefmark{2}sarbani1234@gmail.com},{\IEEEauthorrefmark{3}chinmoy.kundu@ucd.ie},
     {\IEEEauthorrefmark{5}mark.flanagan@ieee.org}    
    \vspace{-0.6cm} }
     
\maketitle
 \thispagestyle{empty}

\begin{abstract}

In this paper, we propose an opportunistic user scheduling scheme in a multi-user reconfigurable intelligent surface (RIS) aided wireless system to improve secrecy. 
We derive the secrecy outage probability (SOP) and its asymptotic expression in  approximate closed form.  The asymptotic analysis shows that the SOP does not depend on the transmitter-to-RIS distance and saturates to a fixed value depending on the ratio of the path loss of the RIS-to-destination and RIS-to-eavesdropper links and the number of users at high signal-to-noise ratio. 
It is shown that increasing  the number of RIS elements leads to an exponential decrease in the SOP.
We also compare our scheme with that of a non-orthogonal multiple access (NOMA) scheduling scheme, which chooses a pair of users to schedule in each time slot. The comparison shows that the SOP of all of the NOMA users is compromised, and that our proposed scheduling scheme has better performance.
\end{abstract}

\begin{IEEEkeywords} 
Asymptotic analysis, opportunistic user scheduling, RIS, and secrecy outage probability. 
\end{IEEEkeywords}

 \section{Introduction}

Reconfigurable intelligent surfaces (RISs) represent a promising solution towards achieving both spectrum and energy-efficient communication with low hardware cost \cite{Zhang_RIS_infty_CLT}.
An RIS can intelligently control the signal propagation by inducing certain phase shifts in the signal to enhance the communication performance \cite{Rui_RIS_dir_opt_singleusr}.

Like any other wireless communication systems, RIS-aided systems are also vulnerable to eavesdropping because of their broadcast nature \cite{poor_Wireless_physical_layer_security}.
The secrecy outage probability (SOP) of an RIS-aided system with a transmitter, a single user, and an eavesdropper was first evaluated in \cite{Renzo_Secrecy_Performance_Analysis}.
A computable SOP expression for an RIS-aided system consisting of a single user and an eavesdropper was derived for vehicular communication in \cite{Ai_Ottersten_Secure_vehicular_communications}.
A computable SOP expression was also derived for an RIS-aided system with a single user and an eavesdropper, assuming discrete RIS phase shifts, in  \cite{Trigui_Zhu_Secrecy_Outage_Probability}.
In \cite{Wang_Ni_Secrecy_performance_analysis}, the SOP of an RIS-aided unmanned aerial vehicle system was evaluated in the presence of a single user and multiple eavesdroppers. 
The secrecy of an RIS-aided system with a single user and an eavesdropper was improved by introducing artificial noise from a jammer in \cite{wafai_gc}. The SOP of the system was derived and then used to implement  optimal power allocation. Although the SOP was evaluated in \cite{Renzo_Secrecy_Performance_Analysis, Wang_Ni_Secrecy_performance_analysis, Trigui_Zhu_Secrecy_Outage_Probability, Ai_Ottersten_Secure_vehicular_communications, wafai_gc}, the articles considered a single-user system. 
 Furthermore,  the expressions in \cite{Ai_Ottersten_Secure_vehicular_communications, Trigui_Zhu_Secrecy_Outage_Probability,  Wang_Ni_Secrecy_performance_analysis} were presented in the form of  Fox H-function, Meijer G-function, and infinite series, respectively, from which it is difficult to gain design insights.

Towards improving secrecy in low-complexity multi-user communication networks, opportunistic user scheduling (OUS) is a  low-complexity and inexpensive  technique to implement and has been extensively studied in the non-RIS aided system context \cite{kundu_dual_hop_regenerative, kundu_Ergodic_secrecy_rate_of_optimal}.  The OUS scheme  involves selecting the best user among all users in each time slot for transmission depending on the channel state information (CSI) of the available links \cite{kundu_dual_hop_regenerative}. 
{The OUS scheme in a multi-user RIS-aided systems was studied in \cite{Zhong_perfana_usersel}, however without secrecy constraints. 
} Opportunistic
user scheduling allows us to achieve the highest achievable rate in each time slot by implementing RIS phase alignment toward the best user without employing complex RIS phase optimization.  
{Phase alignment toward the best user automatically ensures the misalignment of phases toward eavesdroppers, which improves security. 
Furthermore, alignment of the phases of the RIS elements towards each user simultaneously is difficult when a non-orthogonal multiple access (NOMA) transmission scheme is employed \cite{Ding_two_user_NOMA,noma_secrecy_ISR_MIMO, NOMA_secrecy_RIS_GC_21}, which may compromise the individual achievable secrecy rate of some users.}

Opportunistic user scheduling has not yet been investigated for improving secrecy in RIS-aided multi-user communications. 
{Moreover, apart from \cite{wafai_gc}, the articles cited above do not provide any  insights regarding how the system parameters, such as the number of RIS elements and the path loss between the RIS and other nodes, affects  the SOP performance.} 
Motivated by the above discussion,  we propose an opportunistic user scheduling scheme to improve the secrecy of the system 
 of an RIS-assisted system with a transmitter, multiple users, and an eavesdropper.
Our main contributions are listed as follows: 
\begin{enumerate}
    \item  We derive an approximate closed-form expression for the SOP of an OUS scheme where the current user is scheduled by maximizing the user's rate without considering the CSI of the eavesdropping link. 
    \item We obtain an approximate closed-form asymptotic SOP expression which provides useful system design insights, such as the high signal-to-noise ratio (SNR) behavior of the SOP, { the effect of the number of RIS elements on the SOP, and the effect of the pathloss between the RIS and the nodes on the SOP performance. }
    \item {We also compare the SOP performance with that of a NOMA scheduling scheme where a pair of NOMA users is scheduled in each time slot.
    } We show that the SOP of each NOMA user is worse than that of the proposed user scheduling technique.
\end{enumerate}


   \vspace{-0.1cm}
\section{System and Channel 
Model}\label{sec_System_Model}
   \vspace{-0.1cm}
An RIS-aided wireless communication network is considered  that employs an RIS composed of $N$ reflecting elements  between a single-antenna source node S and a cluster of $M$ single-antenna destination nodes $\text{D}^{(m)}$ where $m\in\mathcal{M}=\{1,2,\ldots, M\}$,
in the presence of a single-antenna eavesdropper.
The distance S-R, R-$\text{D}^{(m)}$ for each $m$, and R-$\text{E}$  is $d_{\text{SR}}$, $d_{\text{RD}}$, and $d_{\text{RE}}$, respectively.\footnote{We assume  all R-$\text{D}^{(m)}$ distances are the same. This assumption is adopted for ease of exposition.} We assume that there are no direct links from S to $\text{D}^{(m)}$ for any  $m$ and from S to $\text{E}$ due to deep shadowing.  The link between the node pair $\text{XY}\in\{\text{SR, RE}\}$ corresponding to the $n$-th RIS element where $n\in\{1,2,\ldots,N\}$ is denoted as $h_{\text{XY}}^{(n)}=\sqrt{\zeta_{\text{XY}}}\tilde{h}_{\text{XY}}^{(n)}$ and the links from the $n$-th RIS element to $\text{D}^{(m)}$ for each $n$ and $m$ are denoted as $h_{\text{RD}}^{(n,m)}=\sqrt{\zeta_{\text{RD}}}\tilde{h}_{\text{RD}}^{(n,m)}$, such that $\tilde{h}_{\text{XY}}^{(n)}$ and $\tilde{h}_{\text{RD}}^{(n,m)}$ for each $n$ and $m$ are mutually independent complex Gaussian random variables (RVs) with zero mean and unit variance, and the path loss in dB between any node pair $\textrm{XY}\in\{\textrm{SR}, \textrm{RD}, \textrm{RE}\}$ is  denoted as $\zeta_{\text{XY}}$ (dB) = $z_0-10\upsilon\log_{10}(d_{\text{XY}})$ where $d_{\text{XY}}$ is the distance between the node pair $\text{XY}$, $z_0$ is the reference pathloss in dB,  and $\upsilon$ is the path {loss exponent}. 
We assume that all links experience independent frequency-flat fading. 
The received signal at $\text{D}^{(m)}$ and E through $N$ reflecting elements is written as
\begin{align}
\label{eq_ReceivedSignal_D}
y_{\text{D}}^{(m)}&=\sqrt{P}\sum_{n=1}^N  h_{\text{RD}}^{(n,m)} \exp(j\theta_n) h_{\text{SR}}^{(n)}x+\epsilon_{\text{D}}^{(m)},
\\
\label{eq_ReceivedSignal_E}
y_{\text{E}}&=\sqrt{P}\sum_{n=1}^N  h_{\text{RE}}^{(n)} \exp(j\theta_n) h_{\text{SR}}^{(n)}x+\epsilon_{\text{E}},
\end{align} %
where $P$ is the transmitted energy per symbol, $x$ is the transmitted symbol with unit energy, 
$\theta_n\in(0,2\pi]$ is the adjustable phase shift of the $n$-th RIS element,  $\epsilon_\text{D}^{(m)}$ and $\epsilon_{\text{E}}$ are the additive white Gaussian noise (AWGN) with mean zero and variance $N_0$ at  $\text{D}^{(m)}$ for each $m$ and E, respectively. 


 If the phases of the RIS elements are aligned towards $\textrm{D}^{(m)}$ to maximize its received SNR, i.e., if $\theta_n=-(\phi_n+\psi_{n,m})$ for each $n\in\{1,2,\ldots,N\}$ where $\phi_{n}$ and $\psi_{n,m}$ are the phases of the complex channel $h_{\textrm{SR}}^{(n)}$ and $h_{\textrm{RD}}^{(n,m)}$, respectively,  the received SNR at $\text{D}^{(m)}$ is given by  
\begin{align}
\label{eq_def_gamma_m}
\Gamma_{\text{D}}^{(m)}&=\Gamma_0 \lb({\sum^N_{n=1} \lvert h_{\text{RD}}^{(n,m)}\rvert \lvert h_{\text{SR}}^{(n)}\rvert}\rb)^2,
\end{align} 
 where $\Gamma_0={P}/{N_0}$ is the  transmit SNR.
When $N$ is sufficiently large,  due to the central limit theorem (CLT),  it can be shown that the distribution of the RV within the parenthesis in (\ref{eq_def_gamma_m}) tightly approximates to a Gaussian distribution with mean $\mu_{\textrm{D}}= {N\pi
\sqrt{\zeta_{\text{RD}}\zeta_{\text{SR}}}}/{4}$ and variance $\sigma^2_{\text{D}}=N
\zeta_{\text{RD}}\zeta_{\text{SR}}\lb({16-\pi^2}\rb)/{16}$ and hence,  the distribution of $\Gamma_{\text{D}}^{(m)}$ follows a non-central Chi-squared distribution with a single degree of freedom \cite{Kudathanthirige_Amarasuriya_IRS_rayleigh}, the CDF of which is expressed as \cite{Kudathanthirige_Amarasuriya_IRS_rayleigh}
\begin{align}
\label{eq_cdf_gamma_m_inf}
F_{\Gamma_{\text{D}}^{(m)}}(x)
&=1-\xi Q\Bigg(\frac{\sqrt{\frac{x}{\Gamma_0}}-\mu_{\textrm{D}}}{\sigma_{\text{D}}}\Bigg) ~~\textrm{for~} x\ge0,
\end{align} 
where ${Q}(x)=\frac{1}{\sqrt{2\pi}}\int_x^{\infty}\exp\big(-{t^2}/{2}\big)dt$ is the Gaussian $Q$-function \cite[eq. (2.1.97)]{book_proakis_salehi} and $\xi={1}/{{Q}\big(-\sqrt{\mu_{\textrm{D}}^2/\sigma^2_{\text{D}}}\big)}$ 
is a normalization coefficient to satisfy the PDF constraint.

Due to the specific value of $\theta_n$ chosen for $\textrm{D}^{(m)}$, the channels for the eavesdropper corresponding to $\textrm{D}^{(m)}$ will not be aligned. As a result,   
when $N$ is sufficiently large, the distribution of the effective channel from S to E via R will be complex Gaussian \cite{Renzo_Secrecy_Performance_Analysis}. The received SNR at E corresponding to $\textrm{D}^{(m)}$, $\Gamma_{\text{E}}^{(m)}$, is  exponentially distributed with PDF  
$f_{\Gamma_{\text{E}}^{(m)}}(x)=\frac{1}{\lambda_{\text{E}}}\exp\Big({-\frac{x}{\lambda_{\text{E}}}}\Big)$
and average SNR 
$\lambda_{\text{E}}=N\Gamma_0\zeta_{\text{RE}}\zeta_{\text{SR}}$ \cite{Renzo_Secrecy_Performance_Analysis}. 
In the proposed  OUS scheme, the user among  $\{\text{D}^{(1)},\ldots,\text{D}^{(M)}\}$ whose RIS-assisted channel provides the best user rate without considering the eavesdropping rate is selected to avoid the eavesdropping CSI during scheduling.   
The achievable secrecy rate  for the scheduled user $\text{D}^{(m^*)}$ is
\begin{align}
\label{eq_exact_secrecy}
C_s^{(m^*)}=\max\Big\{\log_2\Big(\frac{1+\Gamma_{\text{D}}^{(m^*)}}{1+\Gamma_{\text{E}}^{(m^*)}}\Big),0\Big\},
\end{align} %
where $\Gamma_{\text{D}}^{(m^*)}=\max_{ m\in\{\mathcal{M}\}}\{\Gamma_{\text{D}}^{(m)}\}$, and 
$\Gamma_{\text{E}}^{(m^*)}=\Gamma_{\text{E}}^{(m)}$ as the SNR for the eavesdropper is independent of any scheduled user. 
The SOP of $\text{D}^{(m^*)}$ is evaluated as   

\begin{align}
\label{eq_sop_no_sel_single}
\mathcal{P}_{\textrm{out}}
&=\int_0^{\infty}F_{\Gamma_{\text{D}}^{(m^*)}}\lb(\rho x+\rho-1\rb)f_{\Gamma_{\text{E}}^{(m)}}(x)dx,
\end{align} 
where $\rho=2^{R_{\textrm{th}}}$ and $R_{\textrm{th}}$ is the required threshold for the secrecy rate. 
\vspace{-0.3cm}
\section{Secrecy Outage Probability}\label{sec_sop}
   \vspace{-0.1cm}
In this section, we derive the SOP of the proposed OUS scheme. The scheduling scheme only takes into account the CSI of the source-to-user links via RIS without requiring the eavesdropping CSI. The SOP of the OUS scheme is provided in Theorem \ref{theorem_sop}.  
{For the derivation of the SOP, we assume (due to the fact that $N$ is large) that $\Gamma_{\text{D}}^{(m)}$ and $\Gamma_{E}^{(m)}$ for each $m$ are independent, even though  the source-to-RIS channel is common among the users and the eavesdropper.} 

\begin{theorem}
\label{theorem_sop}

The approximate SOP of the OUS scheme is
\begin{align}\label{eq_sop_suboptimal_final}
\mathcal{P}_{\textrm{out}}=
\begin{cases}
1-\sum\limits_{m=1}^M \mathcal{V}^{(M,m)}\xi^m J_{+}^{(m)} &\textrm{if~}\mu_{\textrm{D}}^2\Gamma_0\leq \rho-1  \\
1-\sum\limits_{m=1}^M\mathcal{V}^{(M,m)}\xi^m\big(I_+^{(m)}+I_-^{(m)}\big)&\textrm{if~}\mu_{\textrm{D}}^2\Gamma_0>\rho-1 
\end{cases}
\end{align} 
\begin{table*}
\centering
\begin{align}
\label{eq_ss_j_new_theorem}
\mathcal{J}_{+}^{(m,\mathbf{k})}& =\frac{ \Gamma_0}{2\rho\lambda_{\text{E}}\Upsilon^{(m,\mathbf{k})}}
 \Bigg[\exp\Big(-\frac{1}{2(\sigma^{(m,\mathbf{k})}_{\text{D}})^2}\Big(\sqrt{\frac{\rho-1}{\Gamma_0}}-\mu_{\textrm{D}}\Big)^2\Big)+\frac{\mu_{\textrm{D}}\sqrt{\pi}}{(\sigma^{(m,\mathbf{k})}_{\text{D}})^2\sqrt{\Upsilon^{(m,\mathbf{k})}}}\Big. \nn \\  
&\Big. 
\times\exp\Big(\frac{\rho-1}{\rho\lambda_{\text{E}}}-\frac{\mu^2_{\text{D}} \Gamma_0}
{2(\sigma^{(m,\mathbf{k})}_{\text{D}})^2\rho\lambda_{\text{E}}\Upsilon^{(m,\mathbf{k})}}\Big)Q\Big(\sqrt{2\Upsilon^{(m,\mathbf{k})}}
\Big(\sqrt{\frac{\rho-1}{\Gamma_0}}
-\frac{\mu_{\textrm{D}}}{2(\sigma^{(m,\mathbf{k})}_{\text{D}})^2\Upsilon^{(m,\mathbf{k})}}\Big)\Big)\Bigg],\tag{11}  \\
\label{eq_SS_I_closed}
\mathcal{I}_{+}^{(m,\mathbf{k})}&= \frac{ \Gamma_0}{2\rho\lambda_{\text{E}}\Upsilon^{(m,\mathbf{k})}}
\Bigg[\exp\Big(-\frac{\mu^2_{\text{D}} \Gamma_0-\Big(\rho-1\Big)}{\rho\lambda_{\text{E}}}\Big)+\frac{\mu_{\textrm{D}}\sqrt{\pi}}{(\sigma^{(m,\mathbf{k})}_{\text{D}})^2\sqrt{\Upsilon^{(m,\mathbf{k})}}}
\Big.\nn\\&
\Big.
\times\exp\Big(\frac{ \rho-1}{\rho\lambda_{\text{E}}}-\frac{\mu^2_{\text{D}} \Gamma_0}
{2(\sigma^{(m,\mathbf{k})}_{\text{D}})^2\rho\lambda_{\text{E}}\Upsilon^{(m,\mathbf{k})}}\Big)Q\Big(\frac{\sqrt{2}\mu_{\textrm{D}} \Gamma_0}{\rho\lambda_{\text{E}}\sqrt{\Upsilon^{(m,\mathbf{k})}}
}\Big)
\Bigg]\tag{12}. 
\end{align}
\hrule
\end{table*}
where 
\begin{align}
\label{eq_SS_J1}
J_{+}^{(m)}&=\sum_{\mathbf{k} \in \mathcal{S}_m}\binom{m}{\mathbf{k}}
\frac{ w_1^{k_1}w_2^{k_2}w_3^{k_3}}{2^{k_1+k_2+k_3-1}}
\mathcal{J}_{+}^{(m,\mathbf{k})},
\\
\label{eq_SS_I1}
I_{+}^{(m)}&=\sum_{\mathbf{k} \in \mathcal{S}_m}\binom{m}{\mathbf{k}}
\frac{ w_1^{k_1}w_2^{k_2}w_3^{k_3}}{2^{k_1+k_2+k_3-1}}\mathcal{I}_{+}^{(m,\mathbf{k})},
\\
\label{eq_I1_subopt2_new}
I_{-}^{(m)}&=1-e^{-\big(\frac{\mu^2_{\text{D}}\Gamma_0-\lb(\rho-1\rb)}{\rho\lambda_{\text{E}}}\big)}-\sum_{j=1}^{m}\mathcal{V}^{(m,j)}\lb(J_+^{(j)}-I_+^{(j)}\rb),
\end{align}  
with $\mathcal{V}^{(x,y)}=(-1)^{y+1}\binom{x}{y}$, 
$\mathcal{J}_{+}^{(m,\mathbf{k})}$ and $\mathcal{I}_{+}^{(m,\mathbf{k})}$ are given by (\ref{eq_ss_j_new_theorem}) and (\ref{eq_SS_I_closed}), respectively, wherein $\Upsilon^{(m,\mathbf{k})}= \frac{1}{2(\sigma^{(m,\mathbf{k})}_{\text{D}})^2}+\frac{\Gamma_0}{\rho\lambda_{\text{E}}^{(l)}}$, 
$\mathcal{S}_m$ is the set of integer vectors $\mathbf{k}=[k_1, k_2, k_3]$  such that $k_{i}\in\{0 ,\ldots, m\}$ for each $i\in\{1, 2, 3\}$, $\sum_{i=1}^3k_i=m$, 
$\binom{m}{\mathbf{k}}=  \frac{m!}{ k_1!...k_{3}!}$, and $\sigma^{(m,\mathbf{k})}_{\text{D}}= \frac{\sigma_{\text{D}}}{\sqrt{\sum_{i=1}^3k_ip_i}}$.
\end{theorem}
\setcounter{equation}{12} 
\begin{proof}
The evaluation of the SOP in (\ref{eq_sop_no_sel_single}) requires the distribution of destination and eavesdropping channel SNRs. 
{The CDF $F_{\Gamma_{\text{D}}^{(m^*)}}(x)$ is obtained  as}
\begin{align}
\label{eq_SS_cdf}
&F_{\Gamma_{\text{D}}^{(m^*)}}(x)
=\mathbb{P}\lb[\max_{ m\in\{1,\ldots,M\}}\{\Gamma_{\text{D}}^{(m)}\}\le x\rb]\nn\\
&=1-\sum^M_{m=1}(-1)^{m+1}\binom{M}{m}\Bigg[\xi Q\lb(\frac{\sqrt{\frac{x}{\Gamma_0}}-\mu_{\textrm{D}}}{\sigma_{\text{D}}}\rb)\Bigg]^m.
\end{align} 
{The SOP of the system is obtained by substituting (\ref{eq_SS_cdf}) and  $f_{\Gamma_{\text{E}}^{(m)}}(x)$ into (\ref{eq_sop_no_sel_single}) as}
\begin{align}
\label{eq_SS_SOP2}
\mathcal{P}_{\textrm{out}}= 1-\sum^M_{m=1} \mathcal{V}^{(M,m)}     \xi^m \mathcal{P}_{\textrm{out}}^{(m)}.
\end{align} 
where 
\begin{align}
\label{eq_SS_Q}
\mathcal{P}_{\textrm{out}}^{(m)}=\int_0^{\infty}\lb[Q\lb(\frac{\sqrt{\frac{\rho-1+\rho x}{\Gamma_0}}-\mu_{\textrm{D}}}{\sigma_{\text{D}}}\rb)\rb]^m
f_{\Gamma_{\text{E}}^{(m)}}(x)dx.
\end{align} 
The exact solution of the above integral is difficult to derive, hence, we use the following approximation of the $Q$-function from \cite{paper_q_approx}:
\begin{subequations}
  \begin{empheq}[left=Q(x)\approx\empheqlbrace]{align}
  \label{eq_Q_func_sum_3exp1}
    & \sum_{i=1}^3\frac{ w_i}{2}\exp\lb(-\frac{p_i x^2}{2}\rb) ~~~~\textrm{if} ~~x\ge0 \\
  \label{eq_Q_func_sum_3exp2}
    & 1-\sum_{i=1}^3\frac{ w_i}{2}\exp\lb(-\frac{p_i x^2}{2}\rb) \textrm{if} ~~x<0,
  \end{empheq}
\end{subequations}
where
$w_i=\lb\{\frac{1}{6}, \frac{1}{3}, \frac{1}{3}\rb\}$ and $p_i=\lb\{1, 4, \frac{4}{3}\rb\}$ for $i\in\{1,2,3\}$. To apply the approximation, we have to divide the integration regions of $x$ appropriately in (\ref{eq_SS_Q}) 
depending on whether the argument of the $Q$-function is positive or negative in that region. In the region in which the $Q$-function has positive argument, we use (\ref{eq_Q_func_sum_3exp1}) and otherwise we use (\ref{eq_Q_func_sum_3exp2}). 
Taking into account the two cases, and after some trivial manipulations, we obtain  $\mathcal{P}_{\textrm{out}}^{(m)}$ as
\begin{subequations}
\label{eq_SS_Qsolution}
  \begin{empheq}[left={\mathcal{P}_{\textrm{out}}^{(m)}=\empheqlbrace\,}]{align}
&J_+^{(m)} &\quad\textrm{if~}\mu_{\textrm{D}}^2\Gamma_0\leq\rho-1  
\label{eq_SS_Qsolution_a} \\
&I_+^{(m)}+I_-^{(m)} 
&\quad\textrm{if~}\mu_{\textrm{D}}^2\Gamma_0>\rho-1  
\label{eq_SS_Qsolution_b}
\end{empheq}
\end{subequations}
where 
\begin{align}
\label{eq_J1_subopt}
J_{+}^{(m)}&=\int_0^{\infty}\Big(\sum_{i=1}^3\frac{ w_i}{2}\exp\big(\frac{-p_i \chi^2}{2}\big)\Big)^m f_{\Gamma_{\text{E}}^{(m)}}(x)dx,\\
\label{eq_I1_subopt}
I_{+}^{(m)}&=\int_{\alpha}^{\infty}
\Big(\sum_{i=1}^3\frac{ w_i}{2}
\exp\big(\frac{-p_i \chi^2}{2}\big)\Big)^m f_{\Gamma_{\text{E}}^{(m)}}(x),\\
\label{eq_I2_subopt}
I_{-}^{(m)}&=\int_0^{\alpha}  
\Big(1-\sum_{i=1}^3\frac{ w_i}{2}\exp\big(\frac{-p_i \chi^2}{2}\big)\Big)^m f_{\Gamma_{\text{E}}^{(m)}}(x)dx,
\end{align} 
$\alpha=\Big({\mu^2_{\text{D}}\Gamma_0-\lb(\rho-1\rb)}\Big)/{\rho}$, 
and {$\chi=\Big({\sqrt{\frac{\rho-1+\rho x}{\Gamma_0}}-\mu_{\textrm{D}}}\Big)/{\sigma_{\text{D}}}$}. 
{The solutions of $J_{+}^{(m)}$, $I_{+}^{(m)}$, and $I_{-}^{(m)}$ are given by (\ref{eq_SS_J1}), (\ref{eq_SS_I1}), and (\ref{eq_I1_subopt2_new}), respectively. The proofs of these results are provided in Appendix \ref{lemma_Jplus_SS}, \ref{appendix_Iplus_SS}, and \ref{lemma_Iminus_SS}, respectively.}
\end{proof} 
\vspace{-0.1cm}
Although an approximate closed-form expression for the SOP is presented in this section, it is still difficult to obtain insights regarding the effect of the key system parameters $\Gamma_0$, $N$, $M$, $d_{\text{SR}}$, $d_{\text{RD}}$, $d_{\text{RE}}$, $\rho$, and $\xi$. Therefore, in the next section, we will obtain an asymptotic expression of the SOP of OUS scheme.
  \vspace{-0.4cm}
\section{Asymptotic Analysis}
\label{sec_asymptotic} 
  \vspace{-0.1cm}
In this section, we evaluate the SOP in the high-SNR regime where $\Gamma_0 \rightarrow \infty$. 
In addition, we assume $N$ and $M$ are large. 
We make the approximations $1+\Gamma_{\text{D}}^{(m^*)}\approx\Gamma_{\text{D}}^{(m^*)}$ and $1+\Gamma_{\text{E}}^{(m)}\approx\Gamma_{\text{E}}^{(m)}$ as $\Gamma_0 \rightarrow \infty$.
Thus, the SOP from (\ref{eq_sop_no_sel_single}) is approximated  as
\begin{align}
\label{eq_SOP_approx}
\mathcal{P}_{\textrm{out}}&=
\int_0^{\infty}F_{\Gamma_{\text{D}}^{(m^*)}}\lb(\rho x\rb)f_{\Gamma_{\text{E}}^{(m)}}(x)dx. \end{align} 
To find the asymptotic expression, we start with the exact SOP expression derived in Theorem \ref{theorem_sop}.  In the high-SNR regime $\mu_{\textrm{D}}^2\Gamma_0 \gg\rho-1$ and thus the case when $\mu_{\textrm{D}}^2\Gamma_0\leq \rho-1$ in (\ref{eq_sop_suboptimal_final}) does not apply. 
{Furthermore, $\xi\approx 1$  as $N$ is large, which is usually the case in the RIS-aided system.} Hence, due to $\mu_{\textrm{D}}^2\Gamma_0 \gg\rho-1$ and large $N$, we may write the SOP in (\ref{eq_sop_suboptimal_final}) as
\begin{align}
\label{eq_SS_asymp_breakups}
\mathcal{P}_{\textrm{out}}&=1-\mathcal{P}_1-\mathcal{P}_2-\mathcal{P}_3,
\end{align} 
where
$\mathcal{P}_1=\sum^M_{m=1} \mathcal{V}^{(M,m)} \Big(1-e^{-\frac{\mu^2_{\text{D}}\Gamma_0}{\rho\lambda_{\text{E}}}}\Big)$,
$\mathcal{P}_2=\sum^M_{m=1} \mathcal{V}^{(M,m)}  
\sum_{j=1}^{m}\mathcal{V}^{(m,j)}\Big(I_+^{(j)}-J_+^{(j)}\Big)$, and
$\mathcal{P}_3=\sum^M_{m=1} \mathcal{V}^{(M,m)}   I_+^{(m)}$.
It is to be noted that $I_+^{(j)}$,  $J_+^{(j)}$,  and  $I_+^{(m)}$ in $\mathcal{P}_2$ and $\mathcal{P}_3$ are in the high-SNR regime.
Consequently, to obtain $I_+^{(m)}$ in the high-SNR regime, $I_+^{(m)}$ in (\ref{eq_SS_I1}) 
requires $\mathcal{I}_{+}^{(m,\mathbf{k})}$ to be derived in the high-SNR regime. 
Therefore, $\mathcal{I}_{+}^{(m,\mathbf{k})}$  is approximated in the high-SNR regime following Appendix \ref{appendix_Iplus_SS} as 
\begin{align}
\label{eq_Iplus_high_SNR}
\mathcal{I}_{+}^{(m,\mathbf{k})}
&=
\int_{\frac{\mu^2_{\text{D}}\Gamma_0}{\rho}}^{\infty}\exp\Big(-\frac{ \Big({\sqrt{\frac{\rho x}{\Gamma_0}}-\mu_{\textrm{D}}}\Big)^2}{2(\sigma^{(m,\mathbf{k})}_{\text{D}})^2}\Big) f_{\Gamma_{\text{E}}^{(m)}}(x)dx.
\end{align} 
After a change of variable assuming $t = \sqrt{\frac{\rho x}{\Gamma_0}} - \frac{\mu_{\textrm{D}}}{2({\sigma^{(m,\mathbf{k})}_{\text{D}}})^2 
\Upsilon^{(m,\mathbf{k})}}$ in (\ref{eq_Iplus_high_SNR}),  and doing further manipulations, we obtain the solution as
 \begin{align}\label{eq_SS_I_new_asym_lemma8}
&\mathcal{I}_{+}^{(m,\mathbf{k})}
=\frac{\Gamma_0}{2\rho\lambda_{\text{E}}\Upsilon^{(m,\mathbf{k})}}
\Bigg[e^{-\frac{\mu^2_{\text{D}} \Gamma_0}{\rho\lambda_{\text{E}}}} +\frac{\mu_{\textrm{D}}\sqrt{\pi}}{({\sigma^{(m,\mathbf{k})}_{\text{D}}})^2\sqrt{\Upsilon^{(m,\mathbf{k})}}}\nn\\
&\times \exp\lb(-\frac{\mu^2_{\text{D}} \Gamma_0}
{2({\sigma^{(m,\mathbf{k})}_{\text{D}}})^2\rho\lambda_{\text{E}}\Upsilon^{(m,\mathbf{k})}}\rb)Q\Bigg(\frac{\sqrt{2}\mu_{\textrm{D}} \Gamma_0}{\rho\lambda_{\text{E}}\sqrt{\Upsilon^{(m,\mathbf{k})}}
}\Bigg)
\Bigg].
\end{align}
Similarly, we obtain  $J_+^{(j)}$ in $\mathcal{P}_2$ in the high-SNR regime following (\ref{eq_Iplus_high_SNR}) and Appendix \ref{lemma_Jplus_SS}.

Next, after performing some mathematical manipulations, at high-SNR regime, we obtain
$\mathcal{P}_1= 1- \exp({-\frac{\mu^2_{\text{D}}\Gamma_0}{\rho\lambda_{\text{E}}}})$, 
$\mathcal{P}_2=\big(  I_+^{(M)}-J_+^{(M)}\big)$, and 
$\mathcal{P}_3=-\sum^M_{m=1} \mathcal{V}^{(M,m)}I_+^{(m)}$.
We may write $\mathcal{P}_2$ as above due to the fact that  $\sum_{j=m}^{M} \mathcal{V}^{(M,m)}\mathcal{V}^{(m,j)}$ is unity  only if $m=M$ and $j=M$ (zero otherwise). 
We note that as $M$ increases, both $J_+^{(M)}$ and  $I_+^{(M)}$ diminishes in $\mathcal{P}_2$.  As the exponent $m=M$ becomes large in (\ref{eq_J1_subopt}) and (\ref{eq_I1_subopt}), these quantities diminish. 
{In contrast, the magnitude of $\mathcal{P}_3$ increases as $M$ increases.  
} 
Hence, $\mathcal{P}_2$ is negligible as compared to $\mathcal{P}_3$ and thus  $\mathcal{P}_2$ can be ignored when $M$ is large. Therefore, we obtain a simplified asymptotic  SOP expression as
 \begin{align}\label{eq_SS_asymp}
\mathcal{P}_{\textrm{out}}
&=1-\mathcal{P}_1-\mathcal{P}_3
= e^{-\frac{\mu^2_{\text{D}}\Gamma_0}{\rho\lambda_{\text{E}}}}+\sum^M_{m=1} \mathcal{V}^{(M,m)}I_+^{(m)}.
\end{align} 
{ We find from (\ref{eq_SS_asymp}) that the asymptotic SOP consists of two terms. One term is independent of  $M$. The other term which depends on $M$ is negative and increases with increasing $M$. Thus, the OUS scheme improves secrecy. }
 
By substituting  the values of  $\Gamma_0$,  $\lambda_{\text{E}}$, $\mu_{\textrm{D}}$  and $\sigma^{(m,\mathbf{k})}_{\text{D}}$, $I_+^{(m)}$ by (\ref{eq_SS_I1}),  and further performing some mathematical manipulations by replacing the $Q$-function in the expression with its approximation in (\ref{eq_Q_func_sum_3exp1}), we obtain the asymptotic expression in terms of the basic system parameters as
\begin{align}
\label{eq_SS_asymp_final}
&\mathcal{P}_{\textrm{out}}\approx \Bigg(1
 +\sum^M_{m=1}
\frac{W_m\mathcal{A}\frac{\zeta_{\text{RD}}}{\zeta_{\text{RE}}}}
{\lb(\rho+2\mathcal{A}\frac{\zeta_{\text{RD}}}{\zeta_{\text{RE}}}\rb)}
 \Bigg)e^{-\frac{N\pi^2\zeta_{\text{RD}}}{16\rho \zeta_{\text{RE}}}} \nn\\
& +\sum^M_{m=1}\sum_{i=1}^3\frac{W_m \mathcal{A} w_i \pi \frac{\zeta_{\text{RD}}}{\zeta_{\text{RE}}}}
{2\lb(\rho+2\mathcal{A}\frac{\zeta_{\text{RD}}}{\zeta_{\text{RE}}}\rb)}\sqrt{\frac{2\pi\rho N \sum^3_{i=1} k_ip_i}{\lb(16-\pi^2\rb)\lb(\rho+2\mathcal{A}\frac{\zeta_{\text{RD}}}{\zeta_{\text{RE}}}\rb)}}\nn\\
&\times{\exp\Big(\frac{-N\pi^2\frac{\zeta_{\text{RD}}}{\zeta_{\text{RE}} }\lb(\frac{1}{2}+\mathcal{A}p_i\frac{\zeta_{\text{RD}}}{\rho\zeta_{\text{RE}}}\rb)}
{8\lb(\rho+2\mathcal{A}\frac{\zeta_{\text{RD}}}{\zeta_{\text{RE}}}\rb)}\Big)}.
\end{align}
where  
\begin{align}
W_m=\mathcal{V}^{(M,m)}  \sum_{\mathbf{k} \in \mathcal{S}_m}\binom{m}{\mathbf{k}}
\frac{w_1^{k_1}w_2^{k_2}w_3^{k_3}}{2^{k_1+k_2+k_3-1}}
\end{align} 
and  $\mathcal{A}=\Big({\frac{16-\pi^2}{16}}\Big)\big{/}{\sum^3_{i=1} k_ip_i}$. 

 The asymptotic expressions in (\ref{eq_SS_asymp}) and (\ref{eq_SS_asymp_final}) can provide useful insights for designing systems with specific performance requirements by selecting appropriate system parameters.


\textit{Remark 1}:
We observe from (\ref{eq_SS_asymp_final}) that the asymptotic SOP is independent of the values of $P$ and $\xi$, and it saturates to a constant value as $P$ increases.


\textit{Remark 2}:
The asymptotic SOP is not affected by the S-R link path loss  $\zeta_{\text{SR}}$. The SOP also does not depend on the absolute distances between nodes. It only depends on the path loss ratio  ${\zeta_{\text{RD}}}/{\zeta_{\text{RE}}}$. 
{As this ratio increases, the asymptotic SOP decreases. 
}

\textit{Remark 3}:
It is worth pointing that the SOP depends on the product of the factors $\sqrt{N}$ and $\exp(-N)$. As $N$ increases, the factor $\exp(-N)$ dominates, and the SOP decreases exponentially with $N$.

 \vspace{-0.25cm}
 
\section{Numerical Results}
\label{sec_result}
  \vspace{-0.1cm}


This section demonstrates the secrecy performance of the proposed  OUS scheme. Numerical integration (NI) of the analytical integral expressions is also presented to validate the analytical solutions. The simulation results are indicated by red markers, with the same marker as that of the analytical results. Asymptotic  SOP  is shown using dashed horizontal lines.  For numerical results, unless otherwise mentioned we assume $R_{\textrm{th}}=1$ bits per channel use (bpcu),  $(d_{\text{SR}}, d_{\text{RD}}, d_{\text{RE}})=(45, 45, 30)$ meters, $\zeta_0=42$, and $\upsilon=3.5$ following   \cite{Renzo_Secrecy_Performance_Analysis} and \cite{Kudathanthirige_Amarasuriya_IRS_rayleigh}. 
\begin{figure}
\centering
\includegraphics[width=0.45\textwidth]{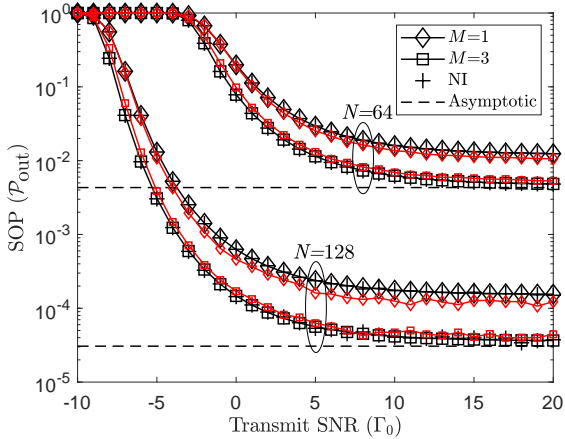}
\vspace{-0.1cm}
\caption{SOP vs. $\Gamma_0$ for the  OUS scheme for different values of $M$  and $N$. 
}
\label{FIG_vary_element_num}
\vspace{-0.6cm}
\end{figure} 

\subsection{Comparison with single-user system }
Fig. \ref{FIG_vary_element_num} plots SOP versus $\Gamma_0$ for the proposed OUS scheme by varying the number of destinations  $M\in\{ 1,3\}$  and the number of RIS elements    $N\in\{64,128\}$.
We notice that the analytical solutions match well with the simulations.
We observe that as the number of users increases, the performance improves significantly by employing the OUS scheme.  
{The performance improvement when $M$ is increased from 1 to 3 is approximately 3 dB  at an SOP of $10^{-3}$ when $N=128$.}  
We notice that at high $\Gamma_0$, the SOP saturates to a constant level, and it no longer improves with an increase in $\Gamma_0$. 
An increase in $N$ also leads to an exponential reduction in the saturation level. 
As $\Gamma_0$ increases, the destination channel SNR improves; however, the eavesdropping channel remains at a fixed disadvantaged position with respect to the destination channel. Hence, saturation occurs. 
This can be verified by examining the ratio of the eavesdropping channel SNR $\lambda_{\text{E}}$ and transmit SNR $\Gamma_0$.



\subsection{Comparison with NOMA scheduling}
In the NOMA scheduling benchmark scheme, we select two users among all the users in each time slot {in our system} as in \cite{Ding_two_user_NOMA} and perform two-user NOMA to transfer data to this user pair {via the RIS}.  
The two-user NOMA scheme is implemented by maximizing the sum rate of the pair \cite{Ding_two_user_NOMA}.
{The NOMA pair is selected in such a way that the channel qualities of the users are at the two extremes to achieve a large performance gain in sum rate over orthogonal multiple-access as shown in  \cite{Ding_two_user_NOMA} and \cite{ding_cooperative_noma_letter}. }
The first user of the NOMA pair is selected to be the best user among all users in exactly the same manner as the proposed OUS scheme (by aligning the phases of the RIS elements towards the user, which maximizes the destination channel rate), and the second user is the worst user, which has the worst destination channel rate given that the phases of the RIS elements are aligned in favor of the the best user. 
A new pair is selected independently at each time slot for scheduling. In this paper, we compare the SOP of the best user and the worst user individually with that of our proposed OUS scheme. As all users have identical channel characteristics, the scheduling schemes are fair to the users in the sense that all the users will be served an equal number of times on average.


\begin{figure}
\centering
\includegraphics[width=0.45\textwidth]{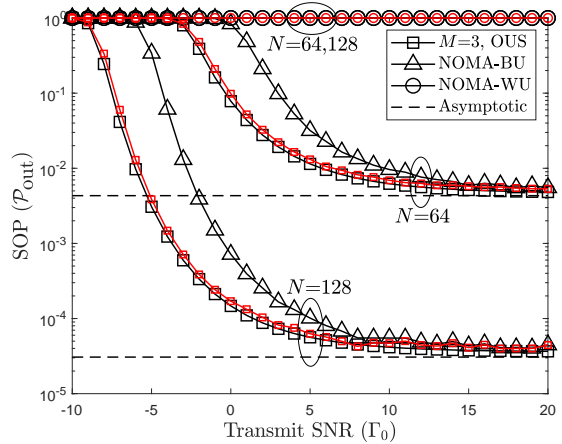}
\vspace{-0.1cm}
\caption{ SOP vs. $\Gamma_0$ for the NOMA-BU, NOMA-WU, and OUS scheme for different values of $M$ and $N$. 
} 
\label{FIG_vary_element_num_NOMA}
\vspace{-0.6cm}
\end{figure} 
Fig. \ref{FIG_vary_element_num_NOMA} compares the SOP of the OUS scheme with that of the NOMA scheduling scheme. The SOP of the best user and the worst user in the NOMA scheme is indicated as NOMA-BU and NOMA-WU, respectively. 
We observe that the SOP of both  NOMA users is worse than that of the OUS scheme. The SOP of NOMA-BU gradually approaches that of the OUS scheme at high SNR, whereas the SOP of NOMA-WU remains extremely low.   
As the best user selection process in the NOMA-BU and OUS schemes is the same, it is expected that the SOP of these schemes merges in the high-SNR regime. As our proposed scheduling scheme opportunistically selects the best user and assigns full allowable power, the performance is better than that  obtained in the NOMA-based scheme. As more power is allocated to alleviate the NOMA-WU's rate by reducing the NOMA-BU's power to improve the sum rate, NOMA BU does not get full allowable power for its transmission. Hence, the SOP of both  NOMA users turns out to be worse than the OUS scheme. In particular, the SOP of NOMA-WU remains extremely poor due to its severely degraded channel condition. We conclude from the observation that the particular NOMA scheduling scheme in this example undermines the security of all the users when compared to the opportunistic scheduling scheme proposed in the paper.    
\section{Conclusion}
An opportunistic user scheduling scheme for an RIS-aided multi-user communication system is proposed to improve secrecy.
The scheme does not require the CSI of the eavesdropping channel.
We derive approximate closed-form expressions of the SOP and its asymptotic limit. Our study provides design guidelines for setting up an appropriate number of users and reflecting elements to achieve a certain level of system performance. 
The analysis shows that the asymptotic SOP saturates to a fixed value depending on the path loss ratio of the RIS-to-destination and RIS-to-eavesdropper links and the number of users.
The asymptotic SOP improves exponentially with the number of RIS elements. 
A performance comparison with a specific NOMA scheduling scheme shows that the proposed scheme outperforms the SOP of the individual NOMA users.  

  \vspace{0.25cm}
\appendices
\section{{Proof of the solution of $J_{+}^{(m)}$ given by (\ref{eq_J1_subopt})}} 
\label{lemma_Jplus_SS}
By applying the multinomial theorem in (\ref{eq_J1_subopt}) we obtain
\begin{align}
\label{eq_I1_prime_subopt}
J_{+}^{(m)}&=\sum_{\mathbf{k} \in \mathcal{S}_m}\binom{m}{\mathbf{k}}
\frac{ w_1^{k_1}w_2^{k_2}w_3^{k_3}}{2^{k_1+k_2+k_3-1}}\mathcal{J}_{+}^{(m,\mathbf{k})},
\end{align} 
where 
\begin{align}
\label{eq_I1_prime_subopt_ext}
\mathcal{J}_{+}^{(m,\mathbf{k})}&=
\int_0^{\infty}\frac{1}{2}\exp\lb(\frac{-\lb(\chi^{(m,\mathbf{k})}\rb)^2}{2}\rb)f_{\Gamma_{E}^{(m)}}(x)dx,
\end{align} 
and $\chi^{(m,\mathbf{k})}=\Big({\sqrt{\frac{\rho-1+\rho x}{\Gamma_0}}-\mu_{\textrm{D}}}\Big)/{ \sigma^{(m,\mathbf{k})}_{\text{D}} }$.
The above equation in (\ref{eq_I1_prime_subopt_ext}) is manipulated by absorbing two $\exp(\cdot)$ functions into one and transforming its argument into a whole-squared form as
\begin{align}
\label{eq_J1_lemma3_aon}
&\mathcal{J}_{+}^{(m,\mathbf{k})}
=\frac{1}{2\lambda_{\text{E}}}\exp\lb(\frac{\rho-1}{\rho\lambda_{\text{E}}}-\frac{\mu^2_{\text{D}}\Gamma_0}{2({\sigma^{(m,\mathbf{k})}_{\text{D}}})^2\rho\lambda_{\text{E}}\Upsilon^{(m,\mathbf{k})}}\rb) \nn \\
&\times\int_{0}^{\infty}
\exp\Big(-\Big(\sqrt{\frac{\rho-1+\rho x}{\Gamma_0}}-\frac{\mu_{\textrm{D}}}{2({\sigma^{(m,\mathbf{k})}_{\text{D}}})^2\Upsilon^{(m,\mathbf{k})}}\Big)^2\nn\\
&\times \Upsilon^{(m,\mathbf{k})}\Big)dx,
\end{align} 
where  $\Upsilon^{(m,\mathbf{k})}= \frac{1}{2(\sigma^{(m,\mathbf{k})}_{\text{D}})^2}+\frac{\Gamma_0}{\rho\lambda_{\text{E}}}$.  After a change of variable assuming $t=\sqrt{\frac{\rho-1+\rho x}{\Gamma_0}}-\frac{\mu_{\textrm{D}}}{2({\sigma^{(m,\mathbf{k})}_{\text{D}}})^2
\Upsilon^{(m,\mathbf{k})}}$ and doing some non-trivial manipulations, we obtain the solution in (\ref{eq_ss_j_new_theorem}).
 \vspace{0.25cm}
\section{{Proof of the solution of $I_+^{(m)}$ given by   (\ref{eq_I1_subopt})}}
\label{appendix_Iplus_SS}
 By applying multinomial theorem in (\ref{eq_I1_subopt}) and doing some further manipulations we obtain
\begin{align}
\label{eq_SS_I1_appendix1}
I_{+}^{(m)}
&=\sum_{\mathbf{k} \in \mathcal{S}_m}\binom{m}{\mathbf{k}}
\frac{ w_1^{k_1}w_2^{k_2}w_3^{k_3}}{2^{k_1+k_2+k_3-1}}\mathcal{I}_{+}^{(m,\mathbf{k})},
\end{align} 
where
\begin{align}
\label{eq_SS_I1_appendix2}
\mathcal{I}_{+}^{(m,\mathbf{k})}
&=
\int_{\alpha}^{\infty}\exp\lb(-\frac{\lb(\chi^{(m,\mathbf{k})}\rb)^2}{2}\rb)f_{\Gamma_{E}^{(m)}}(x)dx.
\end{align} 
We manipulate \eqref{eq_SS_I1_appendix2} as in (\ref{eq_I1_prime_subopt_ext}) 
to obtain the solution in (\ref{eq_SS_I_closed}).

  \vspace{0.15cm}
\section{{Proof of the solution of $I_-^{(m)}$ given by (\ref{eq_I2_subopt})}}
\label{lemma_Iminus_SS}

By applying the binomial theorem in (\ref{eq_I2_subopt}) we obtain 
\begin{align}
&I_{-}^{(m)}
=\int_0^{\alpha}\Bigg[1+
\sum_{j=1}^{m}\mathcal{V}^{(m,j)}\lb(\sum_{i=1}^3\frac{ w_i}{2}
\exp\lb(\frac{-p_i \chi^2}{2}\rb)\rb)^j\Bigg]\nn\\
&\times f_{\Gamma_{E}^{(m)}}(x)dx  
 \end{align}
 The solution of the above integral is obtained in (\ref{eq_I1_subopt2_new}) with the help of (\ref{eq_J1_subopt}) and (\ref{eq_I1_subopt}) utilizing the property of integration.

 \bibliographystyle{IEEEtran}
\bibliography{IEEEabrv,RIS}}
\end{document}